\documentclass[epj,referee]{svjour}
\usepackage{latexsym}
\usepackage{graphics}
\begin{document}
\title{Electronic and magnetic properties of NiS$_{2-x}$Se$_x$: a comparative study}
\author{Cosima Schuster
}                     
\institute{Institut f\"ur Physik, Universit\"at Augsburg, 86135 Augsburg, Germany}
\date{Received: date / Revised version: date}
%
\abstract{
The metal-insulator transition and the problem of
$d$-electron delocalization are investigated in the pyrite system NiS$_{2-x}$Se$_x$ under pressure using 
density functional theory (DFT). We test several approximations, including the generalized gradient approximation (GGA), 
the GGA+$U$ approach, and hybrid functionals.
In addition we apply the $GW$ approximation and perform Hartree-Fock calculations.
The important role of the chalcogen  dimers in the electronic structure is discussed within GGA, which sufficiently
describes the role of the lattice in the metal-insulator transition.
In addition, the magnetic phase diagram is determined. However, the electronic properties are inadequately described --
the insulating ground state of NiS$_2$ cannot be obtained -- and the magnetic order is slightly overestimated.
If correlations are taken into account within GGA+$U$, the insulator is found,
but the non-magnetic ground state of the doped samples is not accessible.
Mixing Fock exchange with local approximations as GGA correctly reproduce the insulating ground state in NiS$_2$,
and metallic ground states with doping and pressure.
The insulator is -- in contrast to earlier suggestions -- of Mott type, i.~e.\, the gap opens between Ni 3$d {e_g}$ states.
Due to the Fock term,  the magnetic order is strongly overestimated.
The applied $GW$ calculations are not able to correct the metallic character of NiS$_2$ and underestimate the gap 
by one order
of magnitude.
}
%
%

\maketitle

\section{Introduction}
Crystal field splitting, electron-electron interaction, and the 3$d$ band width are almost equal in the  pyrites MX$_2$ (M=Fe,Co,Ni,Cu,Zn; X=S,Se)
which gives rise 
to a large variety of electrical, magnetic and optical properties in these compounds. 
Of special interest is the metal-insulator transition  in NiS$_{2}$ with Se-doping \cite{gautier} or under pressure \cite{wilson}. 
The transition is not accompanied by a change in the lattice symmetry and is commonly believed to be driven by 
the electron-electron interaction. 
In addition, a low-temperature antiferromagnetic phase is observed both in the insulating
and metallic regime of NiS$_{2-x}$Se$_x$ \cite{gautier} and by applying pressure in NiS$_2$ \cite{press97}.
The electronic and optical properties of NiS$_2$ were first investigated already 30 years ago \cite{kautz72}, 
and re-investigated experimentally in the late nineties \cite{Mat96,sang97,mam98,prb67,prb68}.  
The renewed interest in transition-metal oxides and their metal-insulator transitions traces back to the attempt to understand the phase diagram of 
the high-T$_c$ superconducting 
perovskites. Moreover, the attempt to use the metal-insulator transition 
of V$_2$O$_3$ or VO$_2$ for technical  applications triggered many studies
up to now \cite{tom09}. 

Theoretical attempts using density functional theory and 
the local density approximation (LDA) to describe the electronic structure of the pyrites date back to
1987 \cite{folkerts}, with the focus on FeS$_2$, where the 
important role of the S(e)-S(e) dimers present in the pyrite structure has been pointed out.
The role of electron-lattice interaction when applying pressure  
was studied ten years later, again in FeS$_2$ \cite{eyert98}.   
Other pyrites are mentioned in \cite{folkerts}, and their electronic structure is explained by filling the band structure 
with the $d$-electrons of the transition metal by going from Fe (3d$^6$4s$^2$ ) to Zn (3d$^{10}$4s$^2$). 
The band insulators are described quite well within the local density approximation. 
however the microscopic origin of the insulating nature of  NiS$_2$ is not well understood. 
Since  more advanced approximations within DFT  are 
available  now, several attempts to improve the material specific theory have been performed.
Recently, Perucchi {\it et al.} \cite{arxiv08} have discussed two different mechanism for the metal-insulator transitions 
based on LDA calculations.  
Compression and  expansion of the lattice alter the metallic state, albeit through two different microscopic mechanisms.
On the other hand, Kune\v{s} {\it et al.} \cite{arxiv09}, using LDA+DMFT, have traced back 
the metal-insulator transition to a single control parameter, namely  the size of the $p$ gap. They confirmed, however, 
that compression and doping influences this control parameter in different ways.
Both studies concentrate on the electronic structure in the paramagnetic phase and neglect the magnetism.

 Hence, we  clarify the role of the S-S dimers and discuss the energy scales mentioned in 
 \cite{arxiv08,arxiv09}.
For the calculations we use the wien2k code \cite{wien2k} and 
the generalized gradient approximation (GGA) in the parameterization  of PBE \cite{pbe}.
 In a next step we determine the 
 magnetic phase diagram, Sec.\ III. Surprisingly, GGA describes the magnetic phase diagram quite well,
 slightly overestimating the phase boundaries as compared to experimental data. 
 In contrast to GGA, LDA always favors the non-magnetic ground state. 
Calculations using improved exchange-correlations functionals are presented in Sec.\ IV.
We present results of GGA+$U$(SIC)  \cite{anisimov} and EECE hybrid \cite{EECE1,EECE2} calculations. To compare with
we have performed also $GW$ calculations on top of  LDA using ABINIT \cite{abinit1,abinitgw}.
We focus on the advantages and shortcomings of the different techniques  with regard to   correlated materials.

\section{The role of the dimers for the electronic structure}

The pyrite structure is a simple cubic lattice with space group Pa$\overline{3}$. It is best described in terms of the
NaCl structure with the transition metal in one sub-lattice and the center of mass of the chalcogenide pairs in the other.
Thus, the S/Se atoms form distorted octahedra around the Ni atoms. In the following, we use the experimental 
lattice constant of NiS$_2$, $a=5.687$ \AA \ \cite{struct_Ni}. 
Applying hydrostatic pressure is simulated by a shortening of the lattice 
constant -- 5 \%/10 \% volume reduction correspond to 1.7 \%/3.5 \% reduction of the lattice constant and to 5/12 GPa, 
respectively. 
In the attempt to understand the phase diagram with doping 
we concentrate on the end points with $x=0$ and $x=2$.
In addition we consider also $x=1$, i.~e. NiSSe. For NiSe$_2$  we also use 
the experimental lattice constant, $a=5.960$ \AA \ \cite{struct_Ni}.
In case of NiSSe, we study two different configurations 
within the pyrite unit cell -- recent STM measurement on doped samples show no sign of a superstructure \cite{Iwaya04} --
and use the theoretically obtained volume, $a_{\rm theo}=5.87$ \AA \, which is about 1 \% larger than 
estimated in experiments \cite{kwi80,gautier}. The first configuration contains S$_2$ and Se$_2$ dimers, 
the second only S--Se pairs. Thereby the second configuration is favored by 0.015 Ryd when the volume is fixed.
In general, S-S, S-Se, and Se-Se pairs are present in doped samples \cite {prb68}. 

Using the mentioned lattice constants the internal positions are obtained by a force minimization.
The distances are summarized in Tab.\ \ref{table_latt}.
In NiS$_2$, the S$_2$ distance is found to be about 2.1 \AA,  
slightly (2 \%) larger than the with experimental data $d=2.07$ \cite{folkerts,prb68}. 
Applying moderate pressure, the dimer decreases only slightly from
2.104 to 2.095, thus the dimer-distance is changed  by 0.4\% when changing  the lattice constant   by 3.5\%. 
Applying even more pressure, corresponding to a 20\% volume effect, it decreases only to 2.07 \AA.
Also when expanding the 
volume drastically, the dimer-distance vary hardly, compare \cite{arxiv09}.
For this reason, 
the metal-insulator transition applying pressure is related to a volume effect, 
since the dimer-distance is nearly less constant when varying the volume.
Turning to the doped samples, we find a large variation of the dimer-distances with doping. 
In NiSe$_2$, the Se-Se distance is given by $d=2.49$ \AA, which is about 4\% larger than the experimental
\cite{folkerts}.
Worth mentioning is that the Se-Se distance is larger than the Ni-Se distance. 
Turning to NiSSe, we observe that in NiSSe I the  
the S-S/Se-Se distances are comparable to dimer-distances of the pristine materials, see Tab.\ ref{tab}. 
In NiSSe II, on the other hand, the S-Se distance corresponds to the averaged value of $d$=2.3\AA. 
\begin{table}
\caption{Lattice constants and typical bond lengths -- given in \AA -- of the systems under consideration, NiS$_2$, compressed NiS$_2$, expanded NiS$_2$, NiSSe, and NiSe$_2$. In addition, the energy differences 
of the bonding and anti-bonding splitting of the chalcogen  (S, Se)
$s$ and $p$ states are given.}  
\label{table_latt}
\begin{tabular}{l|c|cc|}
System& $a$ (\AA)& $d$: Ni--S/Se  & $d$: S(e)--S(e) \\ \hline
NiS2  & 5.28 & 2.21 &2.07 \\
NiS2  & 5.49 & 2.31 & 2.10 \\
NiS2 &  5.69 (exp.) & 2.40   & 2.10\\
NiS2  &  6.04 & 2.56 &  2.11  \\
NiSSe I& 5.87 & 2.48  (S) & 2.14   \\
&  &  2.44 (Se) &   2.42      \\
NiSSe II& 5.87 & 2.41 (S) & 2.30 \\
&  &  2.50 (Se) &  \\
NiSe2 &5.96& 2.48& 2.49 \\
\hline \\
System& $a$ (\AA)& $\Delta E_s$ (eV)& $\Delta E_p$ (eV)\\ \hline
NiS2  & 5.28 & 2.2& 11.0\\
NiS2  & 5.49  &3.2&9.9\\
NiS2 &  5.69 (exp.) & 3.4 & 9.0\\
NiS2  &  6.04 &  2.3& 7.2 \\
NiSSe I& 5.87 & 3.35 (S)& 8.6 (S)  \\
&  &   2.3 (Se)& 8.0 (Se)  \\
NiSSe II& 5.87 & 1.9 &7.7\\
NiSe2 &5.96 &1.8&7.4\\
\end{tabular}
\end{table}

\begin{figure}
\resizebox{0.75\columnwidth}{!}{%
\includegraphics{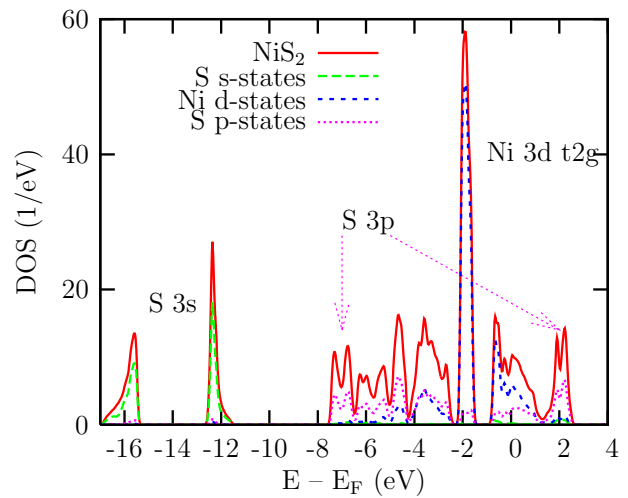}
}
\caption{(Color online) Density of states of NiS$_2$. We show the S 3$s$, S 3$p3$, and Ni 3$d$ states. The bonding and antibonding parts of the S 3$p$ states are indicated.}
\label{fig_dos}
\end{figure}

Next,  we study the volume/dimer-distance effect on the electronic structure, see Fig.\ \ref{fig_dos}, where we show 
the density of states (DOS) of NiS$_2$. 
Generally, the presence of S/Se$_2$ dimers in the pyrites leads to a splitting of the bonding and 
anti-bonding S 3$s$ states as well as S 3$p$ states, as indicated. 
A change of the  volume is expected to modify the bandwidth. 
With increasing volume the bandwidth should decrease, and vice versa. 
However, the situation in NiS$_2$ is more complicated as pointed out for FeS$_2$ in \cite{eyert98}.
Following the detailed analysis in \cite{eyert98} the peaks 
at  $-16$ and $-12$ eV are assigned to S 3$s$ states. 
These states are  observed in the XPS valence spectra at $-15$ and $-11$ eV \cite{sang97}. 
Thus, the splitting is correctly described in GGA, but the position shifted by 1 eV. 
Dominant S 3$p$ contributions 
are found at the lower band edge (at  $-7$ eV)  
and the upper band edge (at 2 eV). They form in itself a double peak structure, compare \cite{eyert98}. 
The  Ni 3$d$ e$_g$ states -- located between $-4.5$ eV  and $-2.54$ eV and between $-1$ eV to 1 eV --
overlap with S 3$p$ states. The  Ni 3$d$ t$_{2g}$ states, located at $-2$ eV, show almost no overlap with the S 3$p$-states.

\begin{figure}
\resizebox{0.75\columnwidth}{!}{%
\includegraphics{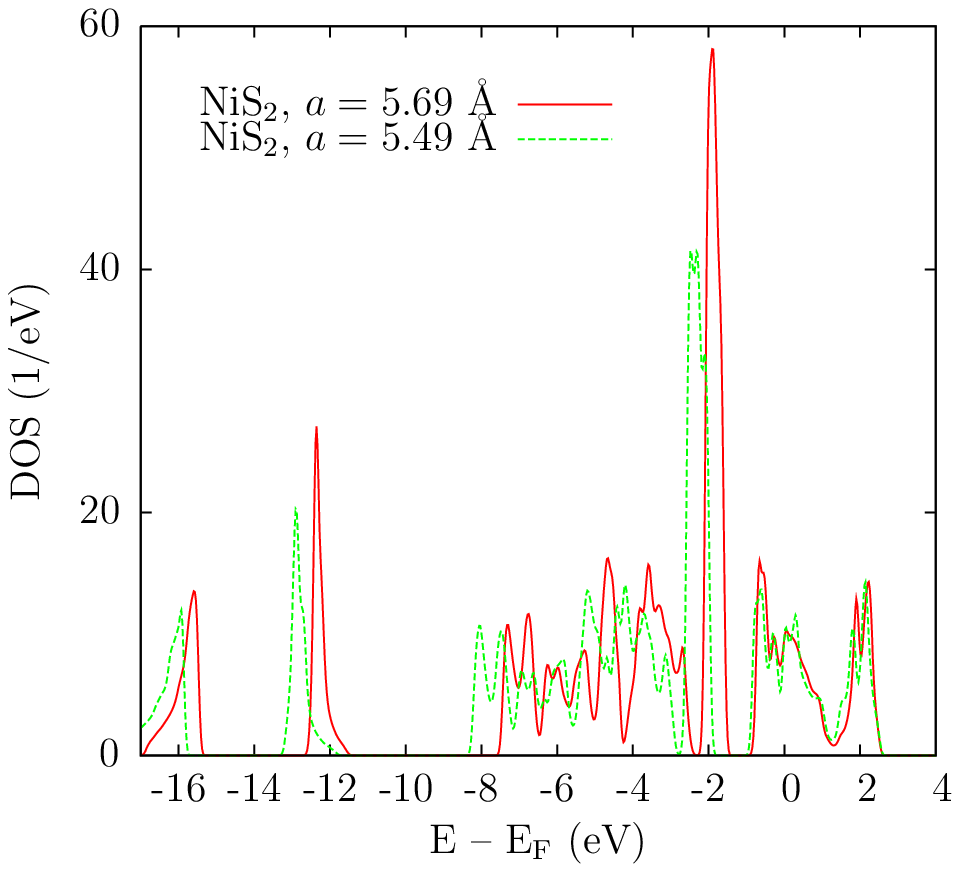}
}
\resizebox{0.75\columnwidth}{!}{%
\includegraphics{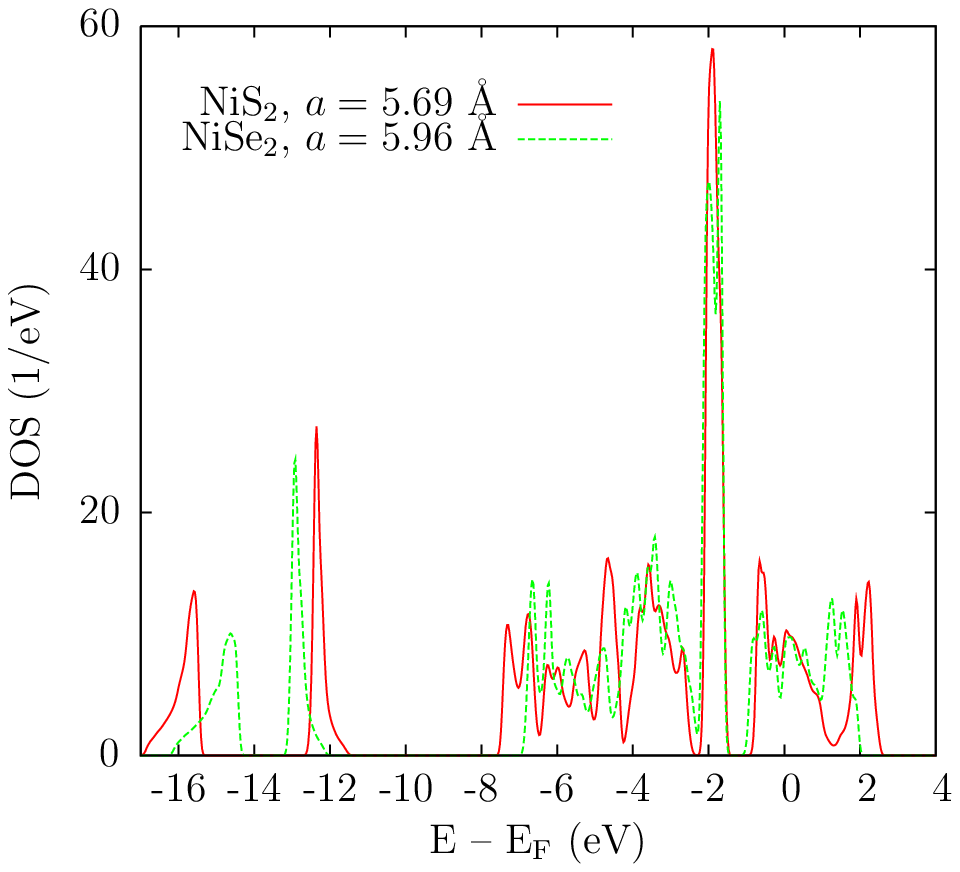}
}
\caption{(Color online) Upper panel: Density of states of NiS$_2$ at the experimental lattice constant and compressed NiS$_2$.
The straight (red) line corresponds to zero pressure, the dotted (green) line to 12 GPa. 
Lower panel: Density of states of NiS$_2$ (red line) in comparison to the density of states of NiSe$_2$ (green line). 
}
\label{fig_dos2}
\end{figure}

Applying pressure, see Fig.\ \ref{fig_dos2}, upper panel, where we show a comparison of the DOS of compressed and uncompressed NiS$_2$, a rigid shift of the $t_{2g}$ by $-0.5$ eV 
due to the enhanced crystal field splitting is observed in the DOS. Due to  this shift the Ni t$_{2g}$- and the S $p$-states overlap. 
The  states below $-3$ eV are subject to an additional shift, 
so that the lower band edge is about 0.6 eV lower for $a=5.49$ \AA than for $a=5.69$ \AA. 
Hence the overall band width is broader and the bonding anti-bonding splitting increases when compressing the lattice. 
In addition to the shift, the binding S 3$p$-states show a slight broadening, but the  
 rigid band shift is the dominant effect.

Upon doping, see Fig.\ \ref{fig_dos2}, lower panel, where we show a a comparison of the DOS of NiS$_2$ and NiSe$_2$, 
the bonding-antibonding splitting is reduced due to the increased Se-Se distance,
both for the $s$  and $p$ states, see Tab.\ \ref{table_latt}, as compared to NiS$_2$.
In NiSSe I, the low energy double peak structure due to the $s$ states splits into four peaks due to the 
presence of both S-S and Se-Se dimers with different dimer lengths.
This could be an experimental signature whether  a structure with both S-S and Se-Se dimers is realized 
or only S-Se pairs are present. The dip at E$_{\rm F}$ is more pronounced in the NiSSe II than in the NiSSe I structure.
In addition, the conduction band width is smaller in structure II (taken at the same volume). 
To summarize, the dominant effect of doping on the DOS/band width is the decrease of the splitting of the S 3$p$-states and the resulting stronger overlap of the Ni 3$d$ with the S $p$ states, as commonly suggested.
The stronger hybridization is identified as the driving mechanism of the metal-insulator transition upon doping. 

\begin{figure}[h]
\resizebox{0.75\columnwidth}{!}{%
\includegraphics{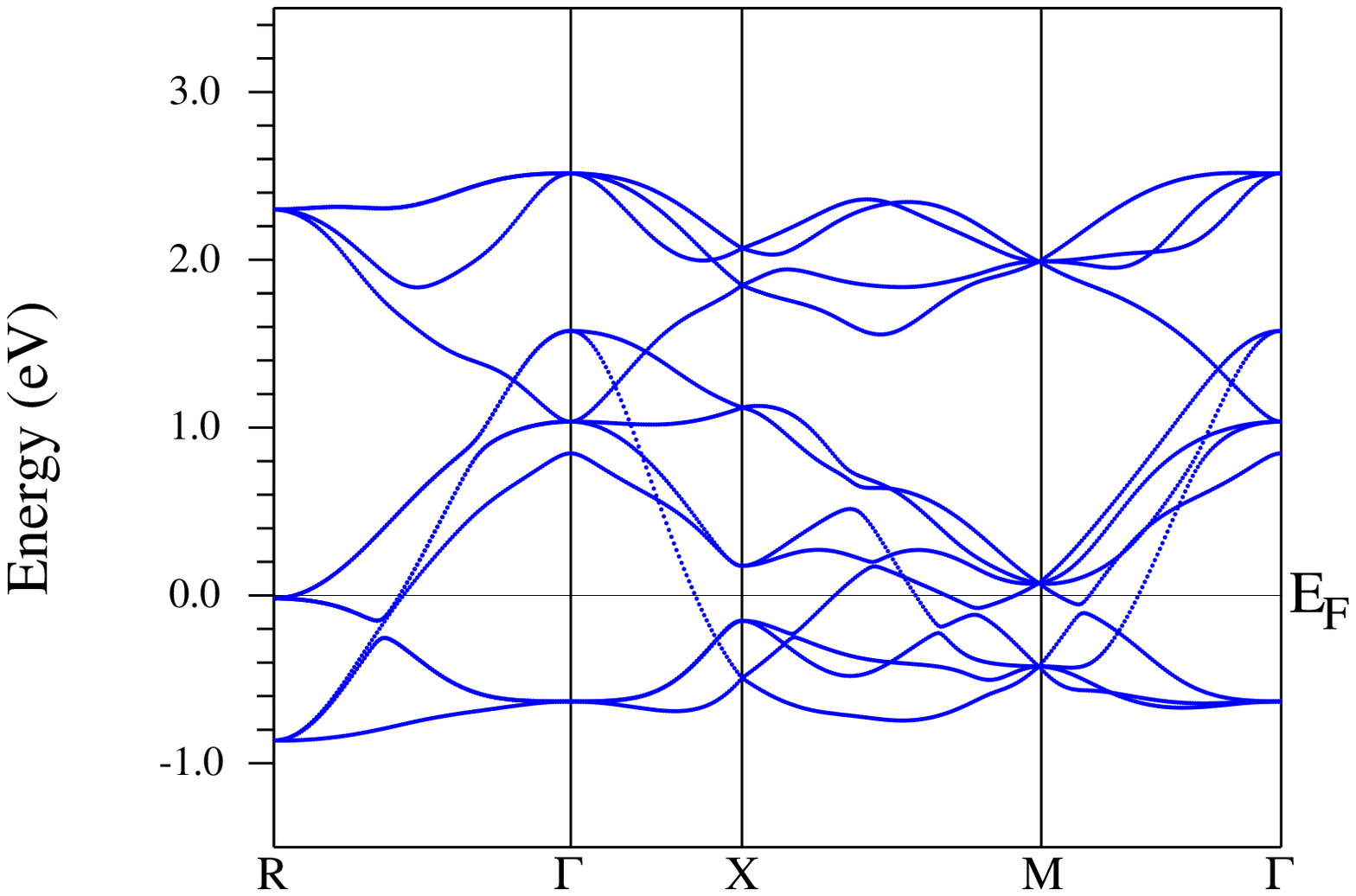} \\
}
\resizebox{0.75\columnwidth}{!}{%
\includegraphics{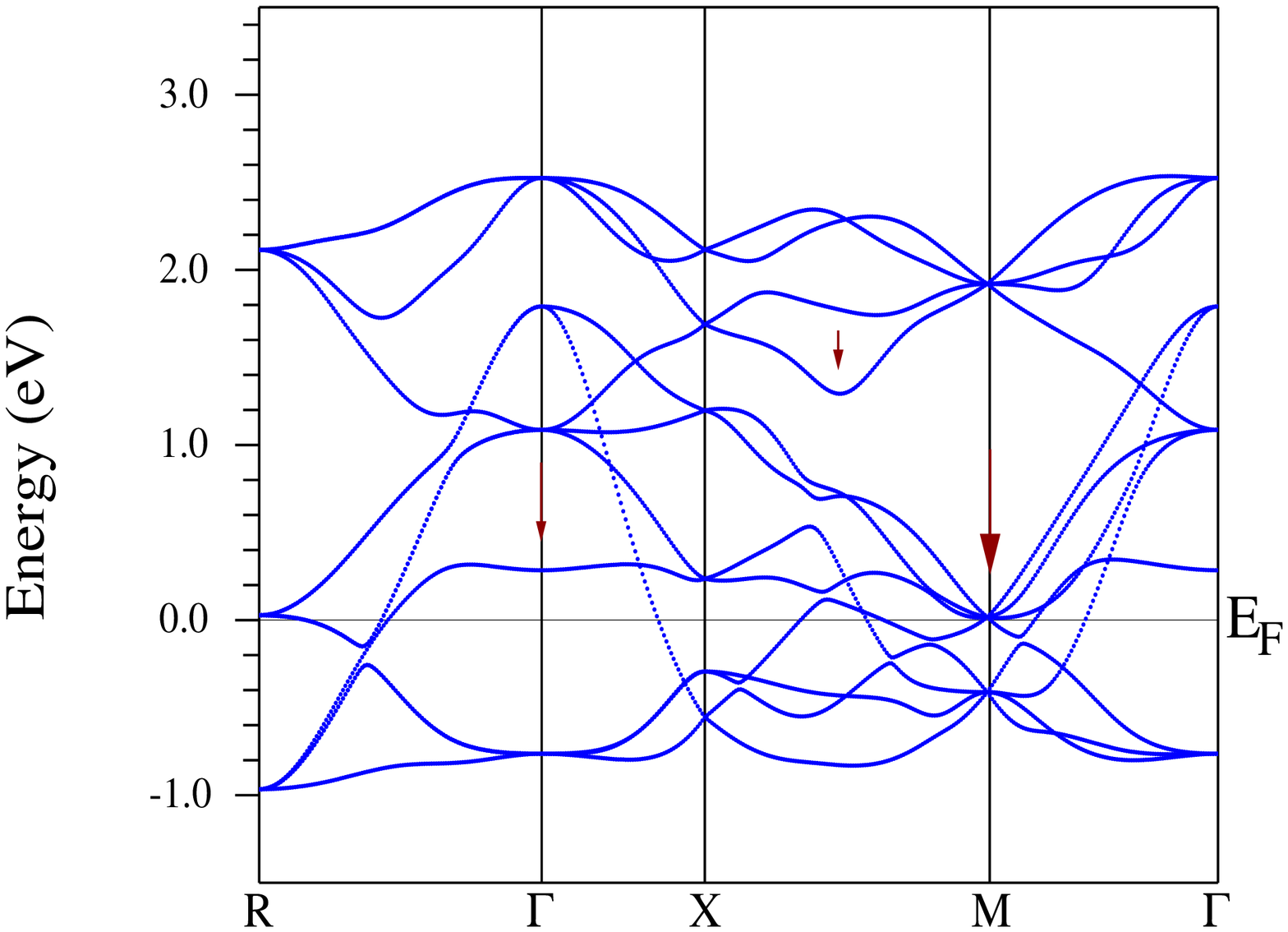} \\
}
\resizebox{0.75\columnwidth}{!}{%
\includegraphics{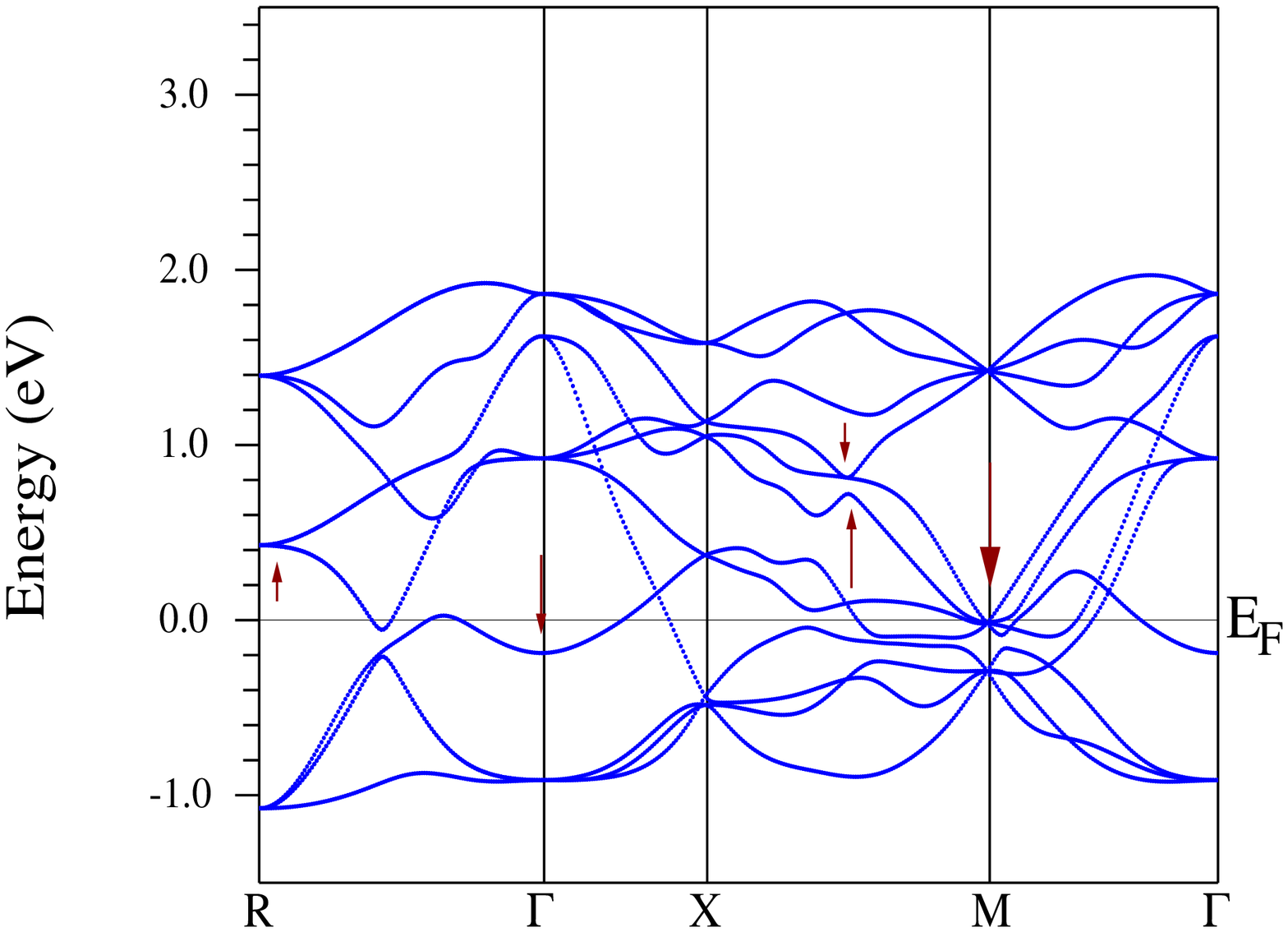}  
}
\caption{(Color online) Band structure near E$_{\rm F}$ of -- from top to bottom -- NiS$_2$, NiS$_2$ with 10\% volume reduction and for NiSe$_2$.}
\label{fig_band2}
\end{figure}

Up to now, we have analyzed the density of states over the whole band width.
Turning to the  underlying band structure,  we take closer look at the states near the Fermi level.
Interestingly, the changes in the dispersion $\epsilon(k)$ around E$_{\rm F}$ 
are similar when applying pressure and with doping.
For comparison, we plot the according band structures  in Fig.\ \ref{fig_band2}. 
The changes in the band structure near the Fermi level with pressure can be focused on two symmetry points:
The band at $\Gamma$ (0.3 eV) moves downwards, almost touching or crossing the Fermi level, respectively.
Second, the four degenerate bands at M  
are shifted towards the Fermi level.  In NiSe$_2$, the band at R moves from the Fermi level to about 0.4 eV above. 
The hybridization due to band shifts or decreased bonding-antibonding splitting does not manifest itself near E$_{\rm F}$.

\section{Magnetic phase diagram}
The phase boundaries of the AFM ordering of type I with pressure and doping can be reproduced within the DFT calculations using GGA.
From the experience with  LDA/GGA studies of other magnetic materials it is expected that LDA underestimates and GGA overestimates magnetic order and magnetic moments, see for example \cite{hafner}. 
In NiS$_2$, we consider an antiferromagnetic alignment of type A, i.~e.\ the magnetic structure consists of 
ferromagnetic planes which are coupled antiferromagnetically. Using this structure no supercell is needed, 
in accordance with \cite{press97}. 
The magnetic properties -- calculated  within GGA -- are summarized in Tab.\ \ref{tab_mag}.
The given magnetic structure is stable within the GGA calculations, frustration effects play a minor role. 
As expected, NiS$_2$ turns out to be a high-spin compound with $\mu=0.73\mu_B$. 
With increasing pressure, the magnetic moment as well as the energy difference to the non-magnetic case decreases. 
Increasing the pressure up to 12  GPa, the antiferromagnetic ground state is still favored, 
but the energy difference to the non-magnetic solution tends to zero.     
NiSe$_2$ turns out to be  non-magnetic, with zero moment. 
At the critical doping (NiSSe), we find also an antiferromagnetic order, which  can be suppressed by applying pressure. 
Unfortunately, a lower boundary of the 
magnetic transition by using LDA could not be determined since LDA favors the non-magnetic solution in all cases. 
This can be related to a volume effect. Within LDA  the ground state volumes are much smaller than the experimental or 
GGA volumes which may have an impact on the spin-polarized calculations. 
For completeness, the data for the energy differences using LDA are also given in Tab.\ \ref{tab_mag}.

\begin{figure}[h]
\resizebox{0.75\columnwidth}{!}{%
\includegraphics{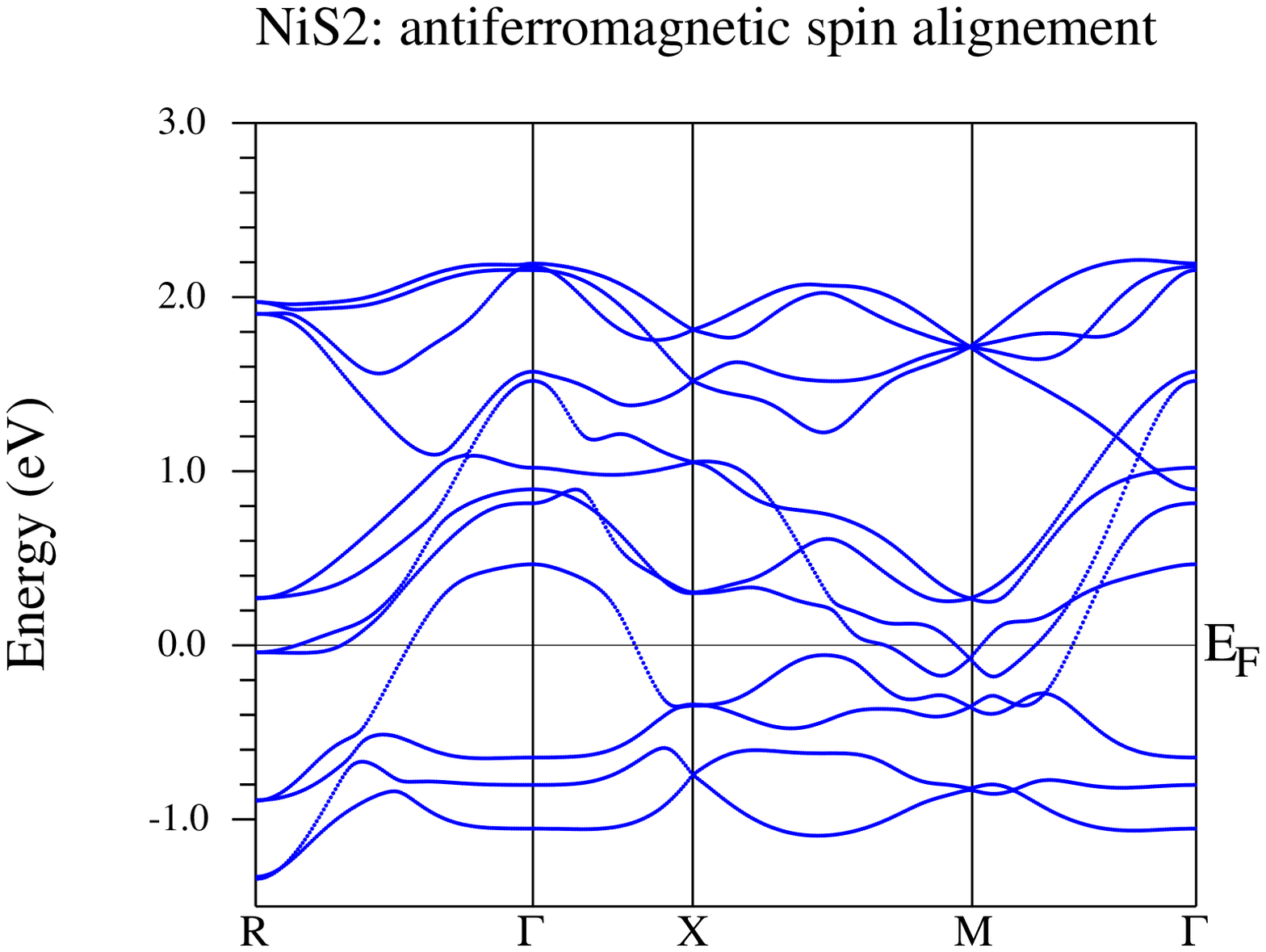}
}
\resizebox{0.75\columnwidth}{!}{%
\includegraphics{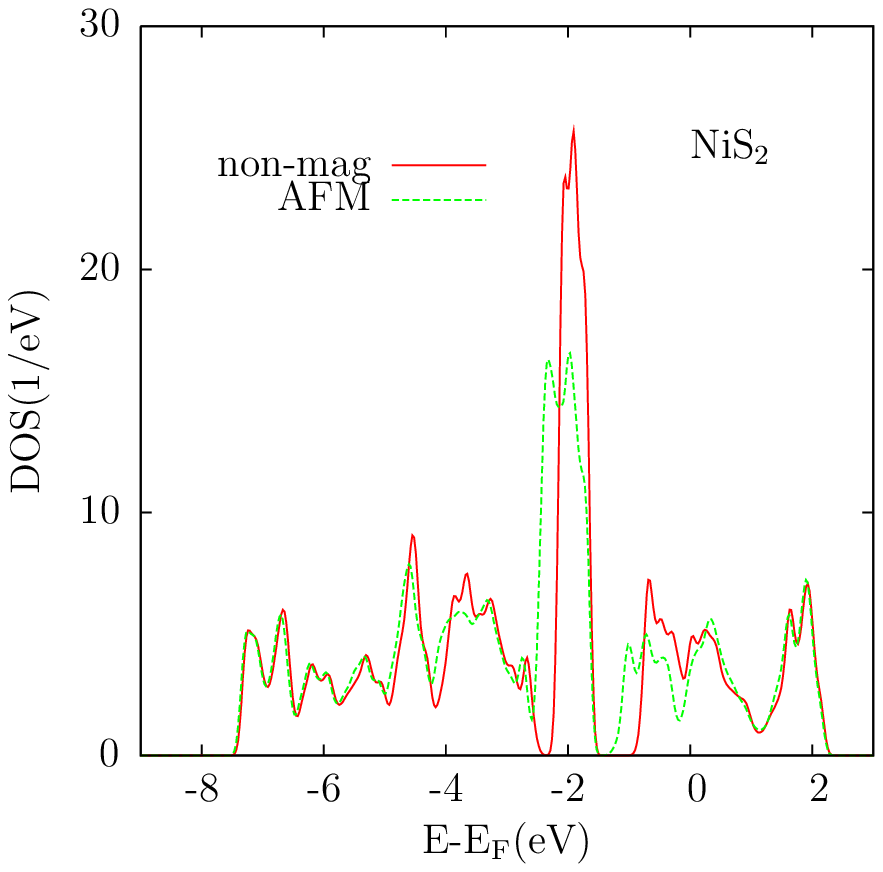}
}
\caption{(Color online) Band structure and density of states of NiS$_2$ with anti-ferromagnetic spin configuration.}
\label{fig_mag}
\end{figure}

The electronic properties change due to the magnetic order.
Band structure and density of states are heavily influenced by the magnetism, see Fig.\ \ref{fig_mag}, where we show DOS and band structure of antiferromagnetic NiS$_2$. 
The degeneracy of the bands is lifted from a fourfold to a twofold degeneracy.
States near E$_{\rm F}$, especially at R and M, are emptied. 
As a result, the dip in the DOS at the Fermi level becomes more pronounced.

\begin{table}
\caption{Magnetic ground state and magnetic moment (in $\mu_B$) within LDA/GGA calculations for the systems under 
investigation. $\Delta E=(E_{\rm non-pol}-E_{\rm mag})$/(number of Ni) is the energy gain due to the magnetic ordering per Ni atom and is given in meV.}
\label{tab_mag}
\begin{tabular}{llcc}
compound & V$_{\rm xc}$ &  mag. moment &  $\Delta E$ \\
NiS$_2$, 5.69 \AA & GGA & $\pm$0.73& 2.1 \\
NiS$_2$, 5.69 \AA& LDA  & $\pm 0.20$ & $-0.14$ \\
NiS$_2$, 5.49  \AA& GGA & $\pm$0.45&  0.17 \\
NiS$_2$, 5.49  \AA& LDA & $\pm$0.11 & $-0.17$\\
NiS$_2$, 5.28  \AA& GGA & 0 & -- \\
NiSSe, 5.69 & GGA &    $\pm$0.4& 0.034     \\
NiSSe, 5.87& GGA & $\pm$0.5 -- $\pm$0.6 & 0.88\\ 
NiSSe, 5.87 & LDA&   $\pm$0.15  & $-0.14$\\ 
NiSe$_2$, 5.96& GGA  & 0& --\\
\end{tabular} 
\end{table}

\section{GGA+$U$, Fock exchange}
Local approximations to DFT like LDA and GGA fail to describe the insulator \cite{arxiv08}.
Recently, a variety of insulating transition-metal oxides have been investigated using LDA+$U$ \cite{anisimov}, 
LDA+DMFT \cite{arxiv09}, Fock exchange or hybrid functionals \cite{EECE2}, or the $GW$ approximation \cite{gattigwvo2}.
In this work, we apply GGA+$U$, hybrid functionals, Hartree-Fock, and  $GW$ calculations to NiS$_2$.
We use the implementation of GGA+$U$, comparing the fully localized limit (SIC) as applied to NiO \cite{anisimov} and the around mean-field treatment  (AFM) \cite{czyzy} for the double counting correction, and the EECE treatment \cite{EECE2} within the wien2k code \cite{wien2k}.

\begin{figure}
\resizebox{0.75\columnwidth}{!}{%
\includegraphics{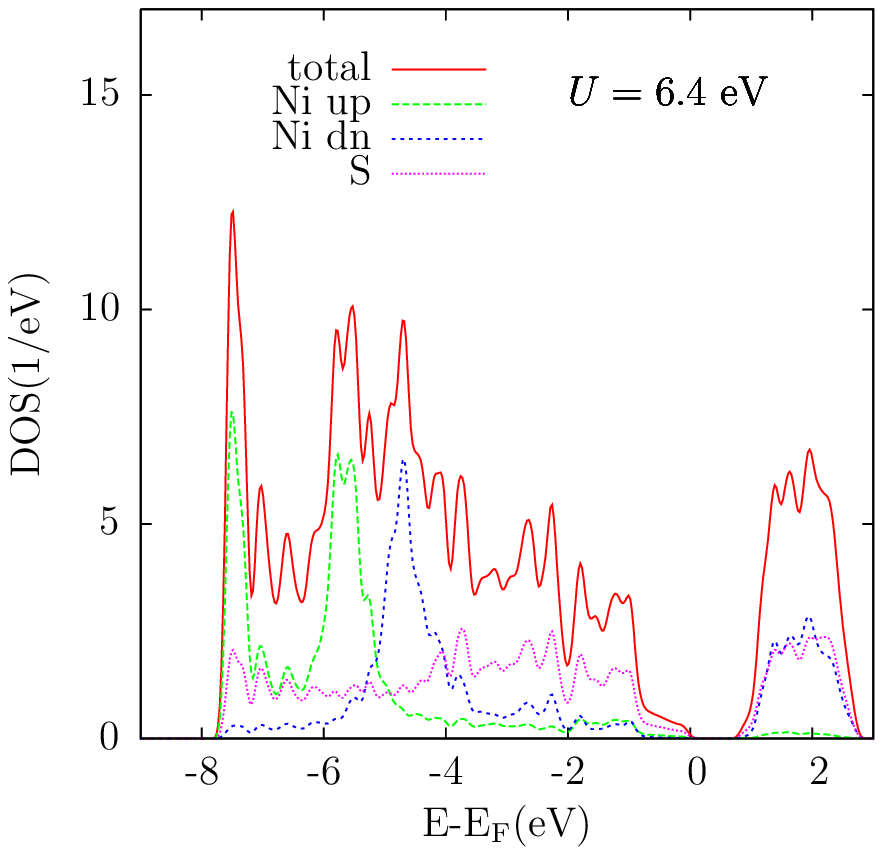}
}
\resizebox{0.75\columnwidth}{!}{%
\includegraphics{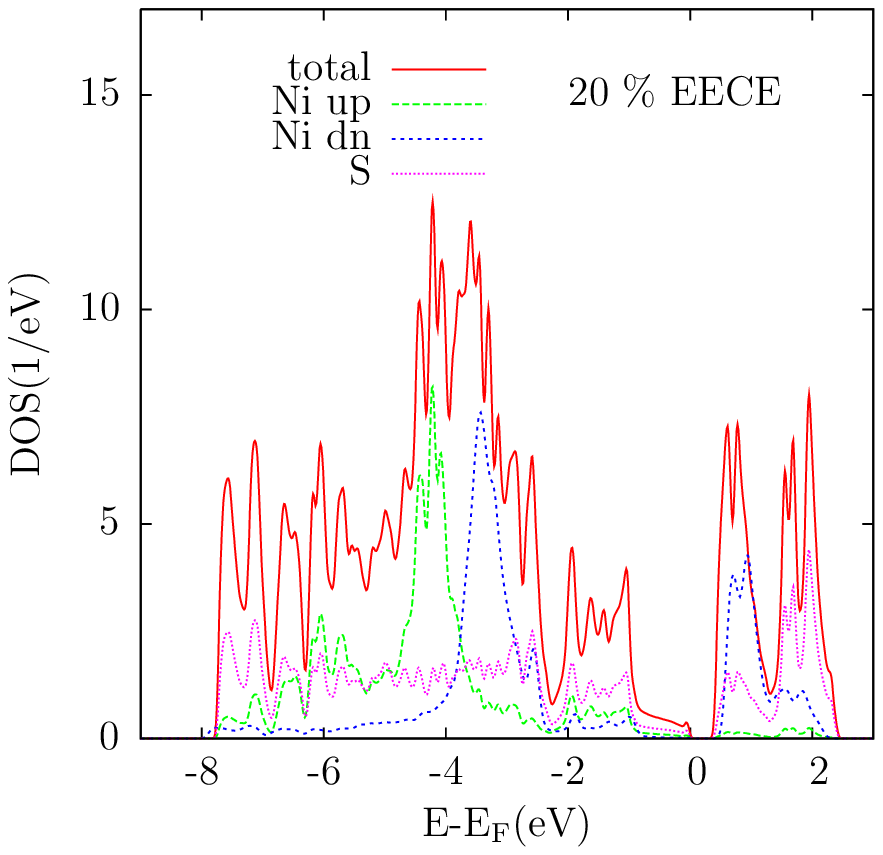}
}
\caption{(Color online) Total density of states of NiS$_2$ (red line) and contributions of the Ni 3$d$ (green and blue line) 
and S 3$p$ (pink line) orbitals: we compare the  GGA+$U$ calculation (left-hand side) with the 
hybrid functional with 20\% EECE (right-hand side). }
\label{fig-DOS-U}
\end{figure}

To start with, we perform constraint GGA calculation to obtain a reasonable value for $U$. 
Following the treatment in \cite{madsenepl} and assuming a Ni d$^8$ configuration as in NiO, 
we obtain $U_{\rm eff}=U-J=6.39$ eV 
for NiS$_2$. $U_{\rm eff}$ decreases to  5.89 eV for NiSSe, and further to 4.94 eV for NiSe$_2$. 
The different chemistry such leads to a considerable decrease of the correlation energy  with doping. 
With decreasing volume,  $U_{\rm eff}$ is constant. 
In the following, we always use the antiferromagnetic spin structure  as discussed in the previous section.
The DOS, calculated with  $U_{\rm eff}=6.4$ eV and the SIC correction, is plotted in Fig.\ \ref{fig-DOS-U}, upper panel,  
and shows a rather poor agreement with 
experimental data \cite{sang97,mam98}. 
The Ni 3$d$ states are shifted to too low energies, 
forming a triple peak structure at the lower band edge. The S $3p$ states
are homogeneously spread over the whole energy range.
Using the experimental value of $U= 5$ eV the gap values, namely $\Delta^{\rm SIC}=0.6$ eV and $\Delta^{\rm AFM}=0.3$ eV,
are in the range of the experimental values, but the  Ni 3$d$ states are located at $-4$ eV instead of $-2$ eV.
A similar behavior was found in case of the 
ferromagnetic metallic pyrite CoS$_2$ \cite{kwon}.

While GGA+$U$ is successful for insulators, 
the description of  correlated metals where the requirement of an integer orbital occupation and well established 
long range order \cite{sawatzky}
is no longer fulfilled may fail. In case of NiS$_2$ under pressure, GGA+$U$  correctly shows that this system is
a non-magnetic metal. 
The results for NiSe$_2$ on the other hand, 
are very sensitive to the choice of the method. Using SIC the ground state is a ferromagnetic metal, in case of AFM the antiferromagnetic metal is nearly degenerate to a ferromagnetic metal with two different moments.

In a next step we perform calculation using Fock exchange for the Ni within the muffin tin spheres (exact exchange for correlated electrons, EECE) and LDA for the correlation potential. This approach is known to give quite reasonable results for correlated electron systems \cite{EECE1,EECE2}.
Instead of a parameter $U$ and the choice for the double counting corrections  which give  deviating results for insulating NiS$_2$ and 
metallic NiSe$_2$,  the mixing parameter of the hybrid functional has to be determined.
As expected 100 \% EECE overestimates the gap by a factor of  5. 
It also shifts the Ni 3$d$ states to the lower band edge, i.~e.\ to $-10$ to $-8$  eV. In this way, 
100 \% EECE is comparable to $U=10.7$ eV.
The gap is induced within the S $p$ states, and not as expected between the Ni $3d$ (Mott-Hubbard type) 
or Ni $3d$ and S $3p$ states (charge transfer type).
If  25 \% EECE is added to LDA the Ni  3$d_{t2g}$ states are found at $-4$ eV. 
The DOS shape corresponds to the DOS obtained for $U=5.4$ eV (suggested $U$ value from \cite{folkerts}). 
Decreasing the amount of EECE further, the gap closes below 20\%; therefore we perform the following calculation with a hybrid functional where 20\% EECE is mixed to LDA. 
Then, gap in the DOS opens within the Ni-S hybridized states,  located between $-2$ and 1 eV,
at the preformed dip in the GGA calculation, as can be seen in Fig.\ \ref{fig-DOS-U}.
The bonding-antibonding S $3p$ states dominate at the lower valence and higher conduction band edge.
The double peak structure of the empty states observed in XAS can be traced back to a Ni 3$d_{eg}$ 
double peak and the S $3p$ antibonding double peak.
Using this hybrid functional, 
an antiferromagnetic solution is obtained, although LDA itself favors the non-magnetic ground state.
The calculated moments become smaller under pressure and with doping but do not vanish.

\begin{figure}
\resizebox{0.75\columnwidth}{!}{%
\includegraphics{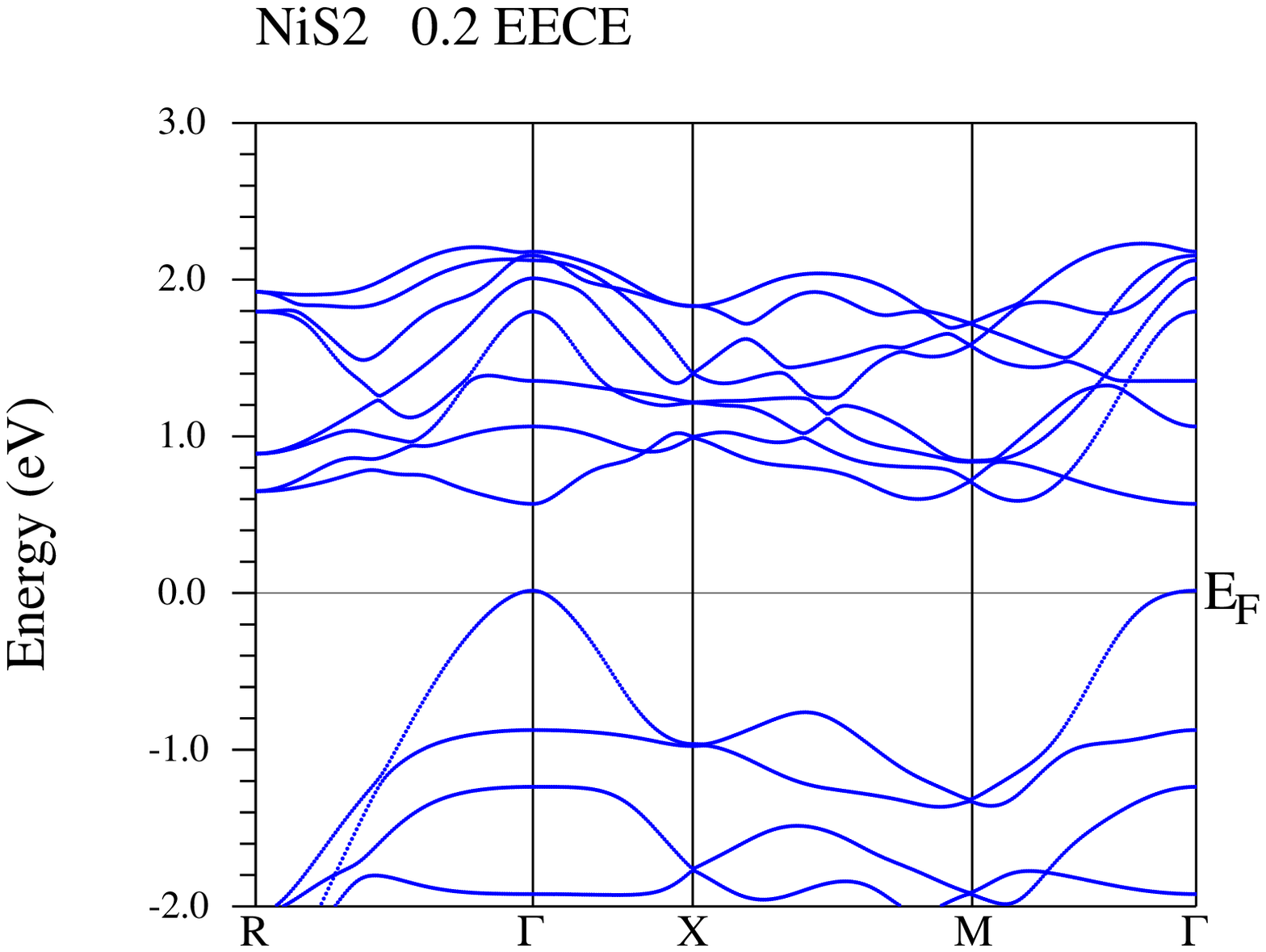}
}
\resizebox{0.75\columnwidth}{!}{%
\includegraphics{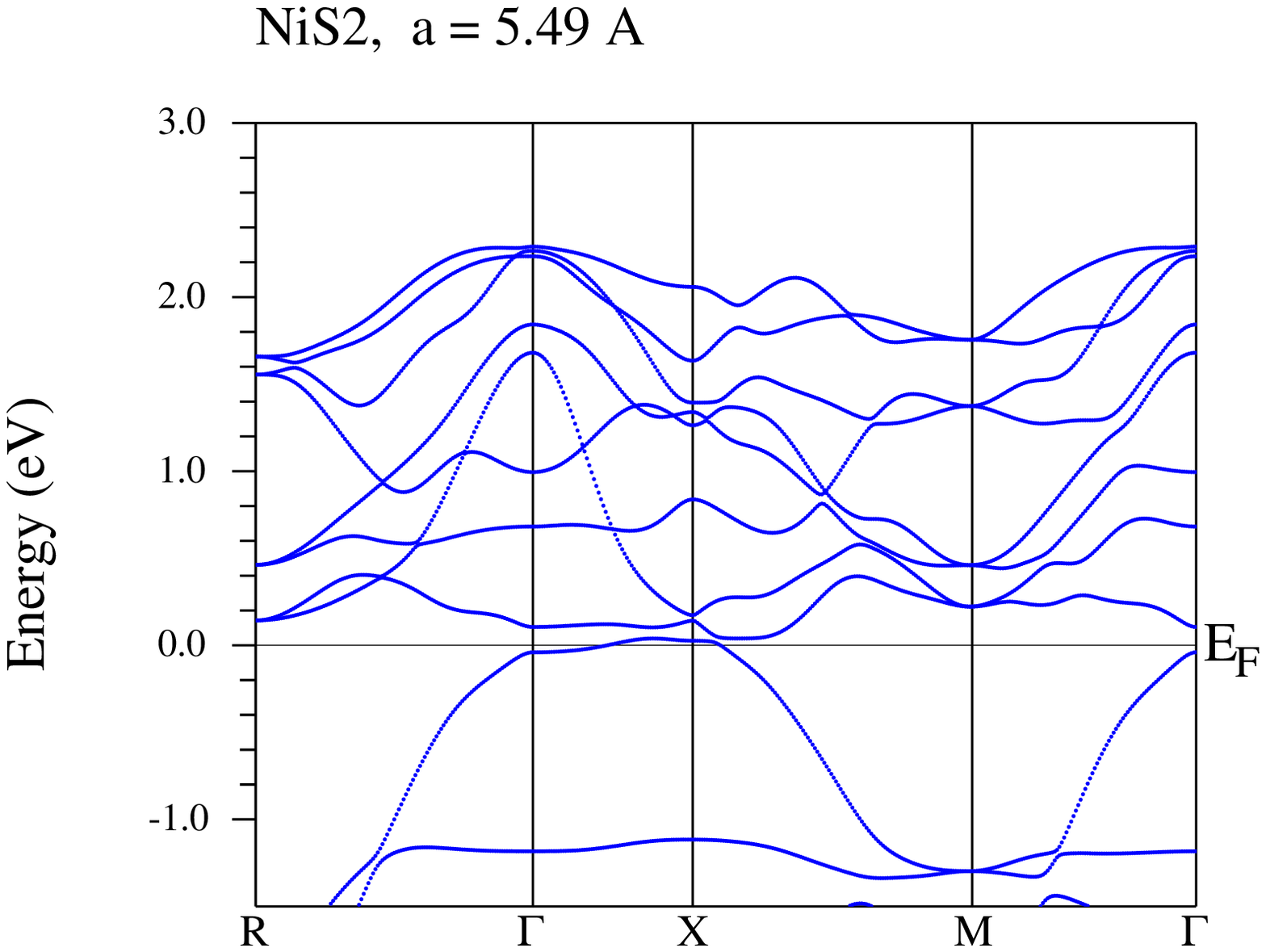}
}
\resizebox{0.75\columnwidth}{!}{%
\includegraphics{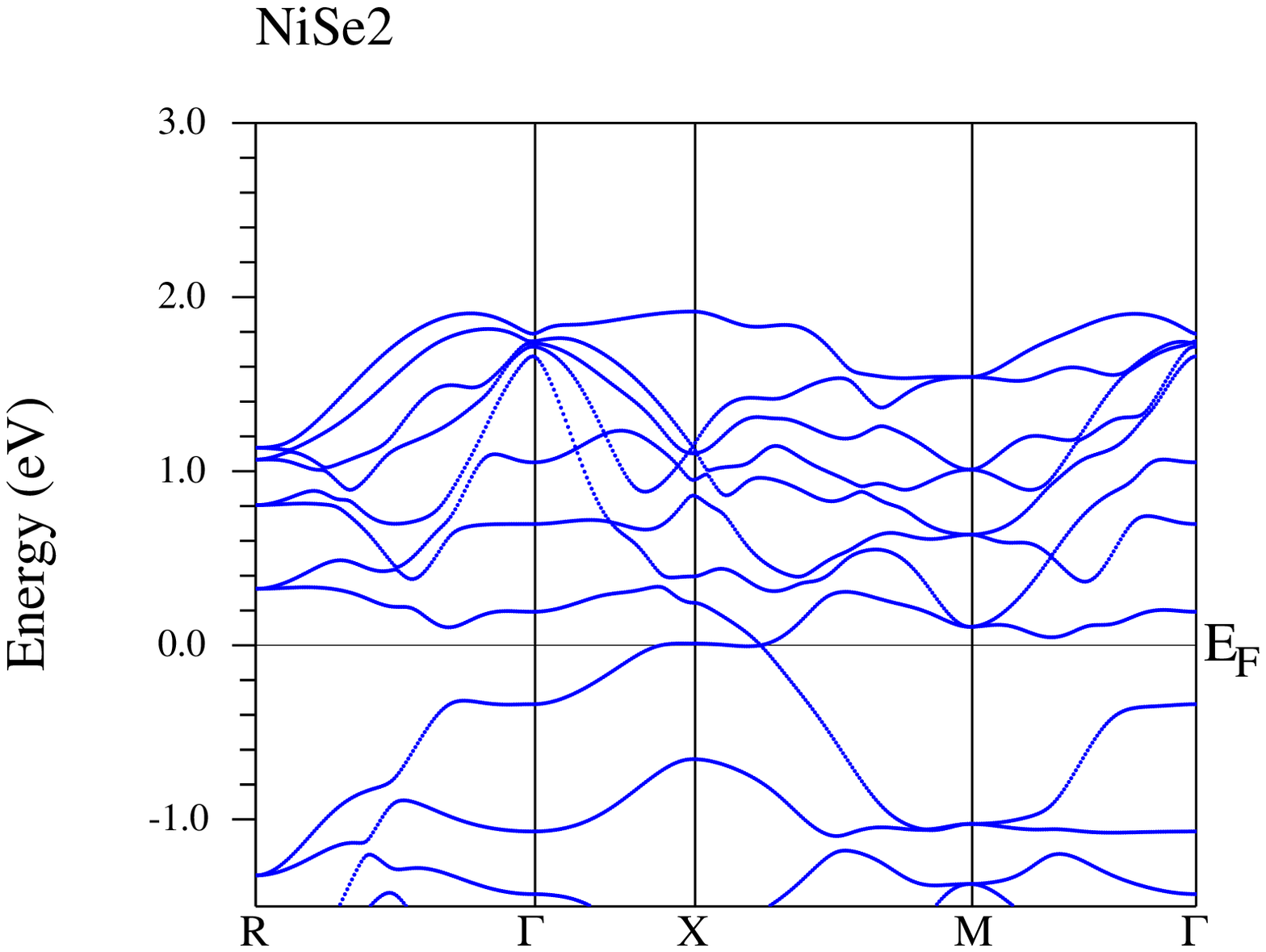}
}
\caption{Band structure obtained using a hybrid functional with 20\% EECE. Top: NiS$_2$; middle: NiS$_2$ under pressure; bottom: NiSe$_2$.}
\label{fig-DOS-U_dop_press}
\end{figure}

The gap of NiS$_2$ is very sensitive to the S-S dimer-distance $d$. 
Using a structure with $d=2.15$ \AA the band gap amounts to 0.3 eV, 
whereas it is 0.6 eV with  $d=2.10$ \AA.  
Increasing the dimer distance to $d=2.36$ \AA, i.e. using the structural parameter of NiSe$_2$ but the lattice constant of
NiS$_2$, the gap closes. 
A closer look at the weighted band structure shows that the gap is located at $\Gamma$; 
the lower band has dominant Ni character, whereas the upper band has dominant S character.
The S character of this band shows up only at $\Gamma$, compare FeS$_2$ \cite{eyert98}.  
According to the band structure the  metal-insulator transition under pressure is triggered by the band
with Ni character which provides the states at the Fermi level between $\Gamma$ and X. 
With doping the metallic state is attributed to the states between X and M.
In this way, a metallic ground state is obtained for NiSe$_2$ due to the increased dimer-distance. 
On the other hand, reducing the  volume also closes the gap at constant dimer-distance. 
The results for the band structure are shown in Fig. \ref{fig-DOS-U_dop_press}, 
where we compare insulating NiS$_2$ with metallic NiS$_2$ under pressure and metallic NiSe$_2$.

To summarize, GGA+$U$ is not able to  describe the system properly, neither with the calculated nor estimated $U$ values.
Hybrid functionals using Fock exchange can reproduce the insulator and give a reasonable DOS.
This approach fails, however,  in describing the magnetic phase diagram by overestimating the ordered phase.
Thus, we have identified a system where the advantage of hybrid functionals versus GGA+$U$ can be clearly demonstrated.

\section{Quasi-particle picture: HF and $GW$}
Corrections to the energy eigenvalues in DFT can be estimated by  first order  many-body perturbation theory with respect 
to the difference in exchange correlation self-energy and exchange correlation potential of DFT. 
Describing the dielectric properties of semiconductors, the $GW$ approximation has become the method of choice \cite{Aulbur}.
Thereby, the non-local, energy-dependent self-energy is approximated  by the Green's function and the screened Coulomb 
interaction, $\Sigma=GW$ 
\cite{hedin}. The screening is calculated on the level of the random phase approximation.
Within the ``one-shot'' procedure $G_0$ and $W_0$ are calculated from LDA/GGA, but 
this $G_0W_0$ approach gives appropriate results only if $GW$ is a small correction to LDA.
Transition-metal oxides can be treated rather well using self-consistent approaches.
For example, the electronic structure of VO$_2$ is described correctly \cite{gattigwvo2}. 
Even  properties of V$_2$O$_3$ can be described \cite{pap09}. 
In NiS$_2$, the electron-lattice coupling is strong and found to be 
the driving mechanism of the metal-insulator transition,
thus it can be expected that the $GW$ approach
is able to  provide a description of the insulating state.
Similar to VO$_2$, in the case of NiS$_2$, a self consistent scheme has to be applied to get rid of the metallic 
starting point. 
In the following we use the $GW$ implementation of the ABINIT package \cite{abinitgw}. 
We compare, see \cite{bruneval}, Hartree-Fock calculations
with  self-consistent, but static $GW$ schemes. 
The self-energy can be written as a sum of the exchange self-energy (static screened exchange), 
the correlation self-energy (Coulomb-hole term) and the dynamical (energy-dependent) self-energy, and the approximations used within $GW$ are classified along this partitioning.
Here, we compare the screened exchange with the screened exchange plus Coulomb-hole.  
Another static $GW$ approximation is the quasi-particle self-consistent $GW$ \cite{faleev}. 
As we start from the Kohn-Sham orbitals obtained within in LDA, 
spin-polarized calculations are in accordance with non-polarized calculations since LDA favors a non-magnetic ground state. 
For simplicity we chose the k-points where the corrections are calculated out of the optimal k-set.  
The given k-points K1 (1/8,1/8,1/8) and K4 (3/8,3/8,3/8) are located between $\Gamma$  and 
R, and K2 (3/8,1/8,1/8) and K3 (3/8,3/8,1/8) between X and M. 
As it can be assumed that the correction do not vary in the vicinity of the chosen k-points,
K1 should give the correction at $\Gamma$, K2 at X, K3 at M, and K4 at R.

On all $GW$ levels, gaps are induced to the band structure of  NiS$_2$ at the chosen k-points, where the smallest gap occurs at K3 (M). 
The results are summarized in Tab.\ \ref{tab-$GW$}.
\begin{table}
 \caption{Energy gaps obtained within different approximation used for the self consistency, namely screened exchange, screened exchange plus Coulomb hole, and model $GW$\cite{faleev}.}
\label{tab-$GW$}
\begin{tabular}{llcc}
method & spin & k-point & gap \\
HF & non-pol & $\Gamma$ & 0.3 eV \\
HF & pol & $\Gamma$ & 0.4 eV    \\
scrEX & pol & M & 0.175 eV  \\
COHSEX & non-pol & M & 0.021 eV  \\
COHSEX & pol & M & 0.025 eV \\ 
QPsc$GW$  & pol & M & 0.027 eV \\
\end{tabular}
\end{table}

The behavior  is in accordance with the GGA calculations as we can assign the dip at E$_{\rm F}$ of the GGA DOS 
to the bands crossing at M.
However, it is found that one band crosses the Fermi level between R and $\Gamma$.
For this reason, $GW$ calculations cannot describe the insulating 
state of Ni$S_2$, even though a gap is formed at all given k-points. 
Therefore, calculations of compressed NiS$_2$ and NiSe$_2$ lead to very similar results for the gaps. 
Moreover, the metallic phases show larger gaps at these k-points, 
but the crossing is present in each case.  

On the other hand, the smallest gap is loacted at  $\Gamma$ within Hartree-Fock, 
in accordance with the results discussed in the previous section. 
In addition, the gap values resulting from the calculations with 20 \% EECE  
 and from the Hartree-Fock calculation are astonishingly of similar magnitude.
However, the total density of states shows no gap at  E$_{\rm F}$ but a large gap between $-5$ to $-1$ eV,  
despite the direct gaps at the chosen $k$-points. 

As a major shortcoming, the $GW$ and Hartree-Fock calculations, 
on top of the  non-magnetic starting point generated by the LDA, 
do not generate a magnetic ground state as the EECE does. The local magnetic minimum is thus not obtained.

\section{Summary}

In summary, the electronic and magnetic structure of doped and compressed NiS$_2$ was discussed in great detail 
within different approximation in the framework of DFT. 
In particular,  LDA/GGA, GGA+$U$, and a hybrid functional composed of 20\% EECE mixed to  LDA have been
compared to Hartree-Fock  and $GW$ calculations.
The metal-insulator transition with applying pressure or doping in NiS$_2$ can be  
described properly  only  when using the hybrid functional. The GGA+$U$ fails to describe correlated metals since is works only well in case of integer orbital occupation and long range order. 
However, as the electron-lattice interaction is very strong in NiS$_2$, and the results very sensitive to changes in the 
lattice structure, the results of LDA/GGA calculations already allows to identify the microscopic origin of the metal-insulator transition, namely the dependence on crystal field splitting and bonding-antibonding splitting.
Additionl corrections to LDA/GGA are then necessary to provide the insulating band structure of NiS$_2$.
 
Therefore, the  magnetic properties of NiS$_2$, on the other hand,  are described best within GGA. 
LDA underestimates the magnetic order, thus no magnetic ground state is found. 
The employed hybrid functional overestimates the magnetic order, even when only a small amount of Fock exchange is mixed 
to LDA. 
The GGA+$U$ approach cannot cpature the  magnetic properties as expected for metallic compounds away from half band filling.

Thus, the Fock exchange applied to the transition metal Ni in the EECE approach gives reasonable results 
as compared to experimental data as far as the electronic properties are concerned
as well as in the insulating as the metallic phases of NiS$_2$. 
The ordering of the magnetic moments is highly overestimated.
The magnetic order present in the EECE approach alters the band structure in comparison to  Hartree-Fock.

\section{Acknowledgements}
We acknowledge fruitful discussions with L. Baldassarre, U. Eckern, V. Eyert, M. Gatti, and J. Kune\v{s}. 
Some of the presented results 
have been obtained through the use of the ABINIT code, a common project of the Universit\' Catholique de Louvain,
Corning Incorporated, and other contributors (URL http://www.abinit.org).
The pseudo-potentials were provided by S. Botti and M. Gatti.
Financial support was provided by the Deutsche Forschungsgemeinschaft (SFB 484)

\end{document}